\documentclass[preprint,showpacs,preprintnumbers,amsmath,amssymb]{revtex4}
\usepackage{graphicx}
\usepackage{dcolumn}
\usepackage{bm}
\begin{document}
\title{
Transport through a multiply connected interacting
meso-system using the Keldysh formalism}
\author{I. V. Dinu$^{1}$, M. \c{T}olea$^{1}$, A. Aldea$^{1,2}$ }
\affiliation{\hskip-1.4cm
$^1$ National Institute of Materials Physics, POB MG-7,
 Bucharest-Magurele, Romania. \\
$^2$ Institute of Theoretical Physics, Cologne University, 50937
Cologne, Germany.}
\date{\today}
\begin{abstract}
We apply the Keldysh  formalism in order to derive a current formula
easy to use for  a system with many sites, one of which is interacting.
The main technical challenge is to deal with the lesser Green function.
 It turns out that,
in the case of the left-right symmetry, the knowledge of the lesser
Green function is not necessary and an exact current formula can be
expressed in terms of retarded Green functions only. The application
is done for a triangular interferometer which gives a good account
of the Fano-Kondo effect. It is found that the interference effects,
in the context of Kondo correlations, give rise to a point in the
parameters space where the conductance is temperature-independent.
We include a comparison with the results from the Ng's ansatz, which
are less accurate, but can be used also in the absence of the above
mentioned symmetry.
\end{abstract}
\maketitle
\section{Introduction}
The transport through  mesoscopic systems has been very
much discussed in the last years because of  promising technological
applications, but also for revealing interesting conceptual aspects.
The simultaneous consideration of both interaction and interference
effects  is a nowadays topic in mesoscopic physics and many efforts
have been done to overcome the specific theoretical difficulties.
 One of the main  tools is the
Keldysh transport formalism, that requires the knowledge of the
retarded and lesser Green functions (e.g. \cite{
meir02,sw,bihary,ueda,moldoveanu}).
The general expression of the current through the lead indexed by
$\alpha$ is :
\begin{equation}
J_{\alpha}=\frac{2e}{h}\int_{-\infty}^{\infty}d\omega
\;i\Gamma_{\alpha}\{f_{\alpha}(\omega)[G^r_{\alpha\alpha}(\omega)-
G^a_{\alpha\alpha}(\omega)]+G^{<}_{\alpha\alpha}(\omega)\}\;,
\end{equation}
where $f_{\alpha}$ is the Fermi distribution function in the
$\alpha$ electrode and $\Gamma_{\alpha}=\pi \tau_{\alpha}^2\rho$,
with the usual notations.

Most of the  papers discuss the Single Impurity Anderson Model (SIAM),
 in which case a simplified version of the formalism can be
employed. In this model, the two leads are connected to the same
site, which mimics the quantum dot, and in this situation the
knowledge of the retarded Green function is sufficient for
describing the transport properties (as the lesser
Green function is eliminated by current symmetrization)\cite{meir02}.
Only a few papers go beyond this model by introducing in the Hamiltonian
a term which produces a short-cut of the impurity,  mimicking a mesoscopic
ring \cite{bulka, hoff}.
In the non-interacting case, eq.(1) can be shown to give the
Landauer-B\"uttiker result, that was widely employed.

The implementation of the Keldysh formalism for the general case of
a many-site, multiple connected system with interaction is not
trivial.
Second order perturbative calculations have been applied for the case of the
electron-phonon interaction\cite{bihary,ueda} or in the context of dephasing
 \cite{moldoveanu}, but such  perturbative approaches cannot be applied
in the case of strong electronic correlations. Recently,
Entin-Wohlman et al. \cite{meir05} and Kashcheyevs et al. \cite{K}
developed an equation-of-motion solution for $G^{r}$ in any complex
geometry described by a tight-binding model, but not for $G^{<}$.
The calculation of $G^{<}$  in the strong interaction regime is a
more difficult task and requires supplementary decoupling
approximations \cite{sw, bulka2004}.

A solution to this problem was  also suggested
 \cite{ew05,ding} by diagonalizing the interacting Hamiltonian
in the slave-boson representation. Then the knowledge
of the retarded Green function is sufficient again,
but the approach restricts the range of validity of the transport
calculation as the mixed valence regime is not correctly described.
Another way to get rid of $G^<$ was proposed in \cite{sun} by imposing
stationarity conditions on the mean values of some non-hermitic
operators, the meaning of which is not obvious to us (especially under
non-equilibrium conditions).

In this paper we propose two different approaches. One is to
approximate the lesser Green functions by using the Ng's ansatz
\cite{ng}. Another possibility is to write down the system of
equations of motion for $G_{ij}^{<}$ (where i,j index any site of
the system)
 and express the current (1) in terms of
$G_{dd}^{<}$ and $G_{dd}^{r}$ (the index 'd' means the interacting site).
It will be shown that,
for the particular case of left-right and time-reversal symmetry,
$G_{dd}^{<}$ can be eliminated by symmetrization of the current
formula. The  result is better in this case than by using Ng's ansatz,
since no  approximations are introduced in the current formula.

 The application is done for the triangle system in the inset of
Fig.1 which is the simplest multiple connected system with two
leads. This system is typical for the study of Fano and Fano-Kondo
effects.

\section{Model and techniques.}
Our approach is based on the following  Hamiltonian written in a
discrete basis :
\begin{equation}\label{a}
\begin{split}
H=\sum\limits_{\textbf{k},\sigma,\alpha}\big(\epsilon_{\textbf{k}\alpha}
-\mu_{\alpha}\big)
c^{\dagger}_{\textbf{k}\alpha,\sigma} c_{\textbf{k}\alpha,\sigma}+
H_{meso}(\{a_{i}^{\dagger},a_{i}\})+H_{T},
\end{split}
\end{equation}
where the mesoscopic system contains an interacting site indexed by $d$:
\begin{equation}\label{b}
\begin{split}
H_{meso}=\sum\limits_{i\sigma}E_i a_{i\sigma}^{\dagger}a_{i\sigma}+
\sum\limits_{ij\sigma} (t_{ij}a_{i\sigma}^{\dagger}a_{j\sigma}+H.c.)
+ H_{int}(a_{d\sigma}^{\dagger},a_{d\sigma})
\end{split}
\end{equation}
and the tunneling term is of the following form
\begin{equation}
\begin{split}
H_{T}=\sum\limits_{\textbf{k},\sigma,\alpha}(\tau_{\alpha}
c_{\textbf{k}\alpha,\sigma}^{\dagger}a_{\alpha,\sigma} + H.c.),
\end{split}
\end{equation}
where $\alpha$ is the lead index (in $c^{\dagger}_{\textbf{k}\alpha,\sigma}$), but also stands for the site where the lead is attached
(in $a_{\alpha,\sigma}$); $\tau_{\alpha}$ is the coupling constant.
 $H_{int}$ may describe any interaction which
implies the site "d", such as Hubbard or electron-phonon (photon)
interaction, the electrostatic coupling to external detectors, etc.
The equations of motion for the lesser Green functions can be written as
\cite{niu}:
\begin{eqnarray}
\sum_{k(\neq
d)}(\delta_{ik}-{g}_{i}^{r}t_{ik})G^<_{kj}=\tau_{\alpha}^2g_
{\alpha}^ < {g}_{i}^{r}G^a_{ij} + {g}_{i}^{r}t_{id}G^<_{dj},~~~ i,j
\ne d.
\end{eqnarray}
In the above equation we use the following notations:
${g}_{i}^{r}=(\omega-E_{i}+i\Gamma_{i})^{-1}$ and $g_{\alpha
}^{<}=2i\pi\rho f_{\alpha }$ is the lesser Green function of the lead
$\alpha$ which is coupled to the site $"i"$; $\rho$ is the flat band
density of states.
The solution  of eq.(5) reads
\begin{equation}
\begin{split}
G^<_{ij}=(A^{-1})_{ik}\big(\tau_{\alpha}^2g_{\alpha }^<{g}_{k}^{r}
G^a_{kj}+{g}_{k}^{r}t_{kd}G^<_{dj}\big),
\end{split}
\end{equation}
where the summation over $"k"$ is assumed, and the notation
$A_{ik}=\delta_{ik}-{g}_{i}^{r}t_{ik}$ have been used.
The function $G_{dj}^<$ in eq.(6) can be obtained in a similar way:
\begin{equation}
\begin{split}
G_{dj}^{<}=\big ((A^{*})^{-1}\big )_{jk}\big(\tau_{\alpha}^{2}g_{\alpha
}^{<}{g}_{k}^{a}G_{dk}^{r}+{g}_{k}^{a}t_{dk}G_{dd}^{<}\big)
\end{split}
\end{equation}
where ${g}_{k}^{a}=({g}_{k}^{r})^{*}$.

One notices that $G_{ij}^{<}$ can be expressed in terms of the
retarded (advanced) Green functions and $G^<_{dd}$. The calculation
of $G^<_{dd}$ still remains a problem and several approximate
solutions can be used, as for instance the Ng's ansatz described
later. However, an important simplification occurs in  the symmetric
two-lead case when, after the symmetrization
$J=(J_{\alpha}-J_{\beta})/2$, the current becomes independent of
$G^<_{dd}$~ :
\begin{equation}\label{general}
\begin{split}
J&=\frac {2e}h \int d\omega\Gamma_\alpha
(f_\alpha-f_\beta)\Bigg\{-\textrm{Im}G^r_{\alpha\alpha}+\\
&+\sum_{\gamma(\neq d)}{(A^{-1})_{\alpha\gamma} {g}_\gamma^r \left [
(-1)^{\delta_{\alpha\gamma}}\Gamma_\gamma G_{\gamma\alpha}^a
+t_{\gamma d} \sum_{\gamma '(\neq d)}(-1)^{\delta_{\alpha\gamma
'}}((A^*)^{-1})_{\alpha\gamma '} \Gamma_{\gamma '} {g}_{\gamma '}^a
G_{d\gamma '}^r \right ] } \Bigg \}.
\end{split}
\end{equation}

The above equation represents the main formal result of this paper.
The current $J$ was expressed solely in terms of different retarded Green
functions, for the symmetric case. The symmetry is necessary in order to get
the same coefficient of $G^<_{dd}$ both in the expression of $J_{\alpha}$ and
$J_{\beta}$, so that $G^<_{dd}$ is eliminated by the symmetrization
 $(J_{\alpha}-J_{\beta})/2$ . One can notice that the SIAM formula is recovered
by the first term in eq.(8). Eq.(8) contains many
retarded Green functions, but, in fact, all of them can be expressed in
terms of $G_{dd}^r$, as  for instance:  $G_{d\gamma}^r=\big (A^{-1}\big
)_{\gamma\beta}g_{\beta}^rt_{d\beta}G_{dd}^r$.

{\bf Ng's ansatz }. The lesser Green function $G^<=G^r\Sigma^<G^a$
is approximated by assuming that $\Sigma^<=\Sigma^{0<} M $ \cite{ng}
where the matrix $M$ is deduced from the relation
$\Sigma^<-\Sigma^>=\Sigma^r-\Sigma^a$ . The result is :
\begin{equation}
\Sigma^<=\Sigma^{0<}(\Sigma^{0r}-\Sigma^{0a})^{-1}(\Sigma^{r}-\Sigma^{a}).
\end{equation}
The intention is again to keep in the final formula only retarded
quantities and  non-interacting functions (trivial to calculate).
The non-interacting selfenergies , for the two lead system read :
\begin{eqnarray}
\Sigma^{0<}=\left(
              \begin{array}{cc}
                f_{\alpha} & 0 \\
                0 & f_{\beta} \\
              \end{array}
            \right)(\Sigma^{0r}-\Sigma^{0a})\\
\Sigma^{0r}-\Sigma^{0a}=-i\left(
              \begin{array}{cc}
                \Gamma_{\alpha} & 0 \\
                0 & \Gamma_{\beta} \\
                 \end{array}
            \right).\nonumber
\end{eqnarray}
It is now straightforward to express the quantities required by the
current formula eq.(1) as :
\begin{eqnarray}
G^<_{\alpha\alpha}&=&-G^r_{\alpha\alpha}f_\alpha[(\Sigma^{r}-\Sigma^{a})G^a]_{\alpha\alpha}
-G^r_{\alpha\beta}f_\beta[(\Sigma^{r}-\Sigma^{a})G^a]_{\beta\alpha}\\
f_\alpha(G^r-G^a)_{\alpha\alpha}&=&f_\alpha[G^r(\Sigma^{r}-\Sigma^{a})G^a]_{\alpha\alpha}
=G^r_{\alpha\alpha}f_\alpha[(\Sigma^{r}-\Sigma^{a})G^a]_{\alpha\alpha}+
G^r_{\alpha\beta}f_\alpha[(\Sigma^{r}-\Sigma^{a})G^a]_{\beta\alpha}.\nonumber
\end{eqnarray}

Finally, the current formula becomes :

\begin{equation}
J_{\alpha}=\frac{2ie}h\int{d\omega(f_\alpha-f_\beta)\Gamma_\alpha
G^r_{\alpha\beta}[(\Sigma^{r}-\Sigma^{a})G^a]_{\beta\alpha}} .
\end{equation}
 The same scheme was used in \cite{Sergueev,Zhang},
but for a different problem, namely the spin transport through a
single-site dot coupled to magnetic leads.

\section{Application and discussions.}
The exact formula eq.(8) will be applied to the particular case of a
triangular interferometer with Hubbard interaction $ H_{int}=
Ua_{d\uparrow}^{\dagger}a_{d\uparrow}a_{d\downarrow}^
{\dagger}a_{d\downarrow}$ .
In the previous section, we reached our goal of expressing the current through
the retarded Green functions only. One important advantage is that there are
already recipes for computing these functions in different approximations.
  We shall use the scheme proposed by Entin-Wohlman {\it et  al}
\cite{meir05} for  $U\rightarrow\infty$. The choice is justified by
the simple analytical formulae which have an easy implementation.

The triangular system  is a good tool for studying
the  interplay between  correlation and interference processes.
 The  non-interacting conductance (curve no.3 in Fig.1)
shows a typical Fano line presenting both a Fano zero
and  a perfect constructive interference.
This is  the result of the interference between the partial waves
passing the dot and the reference arm.

 In the Kondo regime the dot transmits
through the Kondo peak and the transmission phase is "frozen" at
$\pi/2$ (as measured also experimentally
 in \cite{Heiblum,Sato}). The result consists in a much slower variation
of the interference conditions giving rise to a Fano line of reduced
amplitude  compared to the non-interacting case. This is known as the
Fano-Kondo effect and is described by the curve no.1 in Fig.1.

Significant differences can be noticed between the conductance
 obtained by using the exact formula eq.(8) and the Ng's approximation
eq.(12) \cite{note}.
 The differences are rather large, indicating that the Ng's ansatz does
not capture  well the combined effect of correlation and
interference. Fig.1 shows that the destructive interference is
overestimated. When the dot is empty ($E_d \gtrsim 0.1$) all the
three curves coincide, as expected. The main control parameter (also
in experiments) is the position on the  energy scale of the atomic
energy $E_d$ which can be changed by applying an external bias .
Asymptotically, for $|E_d|\rightarrow\infty$, the interacting site
"d" is decoupled from the other sites; consequently the transport is
performed only through the background branch (i.e., that one
connecting the sites "$1$" and "$2$"). The background conductance is
also plotted in Fig.1 and represents the control limit of our
calculations.
\begin{figure}
\includegraphics[scale=0.55]{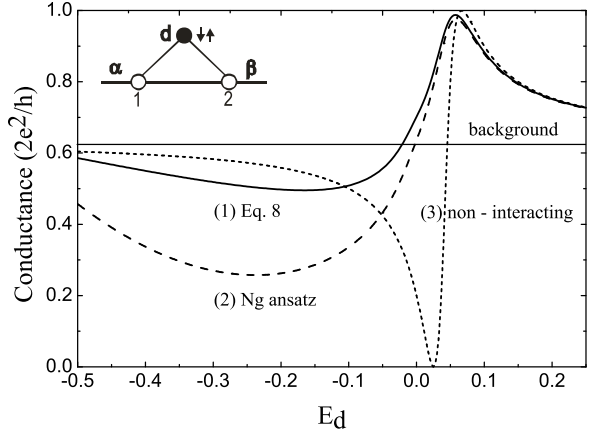}
\caption{Conductance through the triangular interferometer (see
inset): (1) using exact formula (Eq.8), (2) with Ng's ansatz
(Eq.12), and (3) for non-interacting case. The parameters are
$E_1=E_2=E_{Fermi}=0, \Gamma=0.025, t_{12}=t_{1d}=t_{2d}=0.02,
T=10^{-20}$ (measured in units of half-band width). The horizontal
line represents the asymptotic value for $E_d$ going to $\pm\infty$
.}
\end{figure}
\begin{figure}
\includegraphics[scale=0.55]{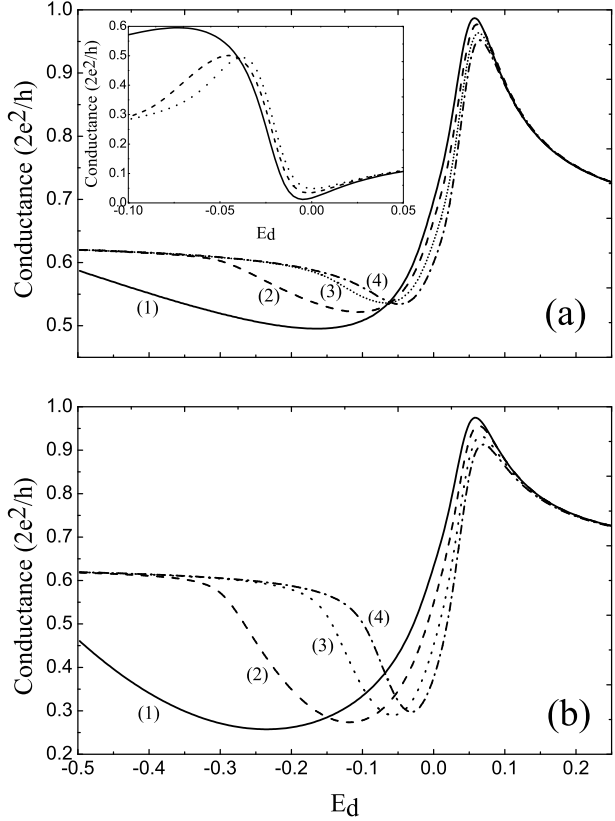}
\caption{a) The crossing point exhibited by the current curves
calculated at different temperatures in the range
 $[T=10^{-20}-10^{-3}$]. Curve (1) is for the lowest temperature.
Other parameters are the same as in Fig.1. The inset shows the same
behavior for the T-shape system. b) The same curves calculated with
Ng's approximation, which misses the crossing point.}
\end{figure}
\begin{figure}
\includegraphics[scale=0.55]{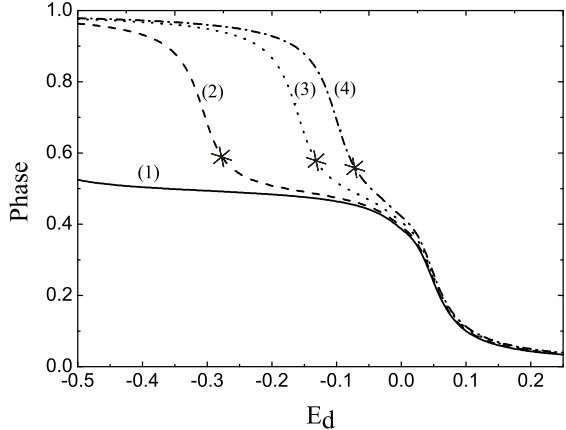}
\caption{The phase of $G_{dd}^{r}$ (in units $\pi$) for the same
parameters as in Fig.2; the stars indicate the gate potentials where
the Kondo temperature equals the plot temperature. The curve (1) is
still in Kondo regime at $E_d=-0.5$.      }
\end{figure}
In the range of the gate potential where the correlations are
important, the temperature dependence is expected to be of Kondo-
type ,i.e. the conductance increases with decreasing temperature.
However, when multiple paths are possible (and interference plays an
important role), an increased transparency of the quantum dot may
give rise, on the contrary, to a reduced conductance of the whole
system. The interference conditions can be changed by a magnetic
field (as in \cite{bulka}) or by variation of the gate potential as
we are doing here.

The isotherms of the conductance $g= dJ/d\mu$  plotted in Fig.2 show
two different temperature regimes separated by a crossing point: on
the right the conductance decreases with $T$, and behaves oppositely
on the left side. Our calculation  determines the crossing point
with an accuracy within numerical errors, and suggests that there is
a gate potential $E_d^c$ such that
\begin {equation}
dg (E_d^c,T)/dT = 0.
\end {equation}
The same crossing point was found for a more simple model, namely
the T-shape system. The T-shape geometry consists of two coupled
quantum dots, only one of them (the non-interacting dot) being
connected to leads. The advantage of this more simple model is that
the conductance is simply expressed by the density of states at the
site connected to leads (details are found in, e.g. \cite{meir05})
and the effect of the interference is more obvious. Basically, the
T-shape and the triangle describe the same Fano-Kondo physics;
the triangle is however  a technical challenge and the first step
versus more realistic models.

The DoS at the Kondo dot always shows the specific Kondo peak at the
Fermi energy. However, the DoS at the coupling site may show a Kondo
peak or dip  depending on the constructive or destructive
interference conditions, respectively. The two cases give opposite
temperature behavior. The interference conditions gradually change
with the applied gate, giving rise to regions with opposite
temperature dependence separated by a crossing point (see inset of
Fig.2).
 The existence of this point seems to be a fingerprint of the
Fano-Kondo effect in systems with interaction and interference.
The use of the exact formula eq.8 is
essential, since  the Ng's approximation eq.12 misses
the crossing point (as can be noticed in Fig.2b). This shows that
the Ng's approximations fails not only quantitatively but also
qualitatively.

If we intend  to identify  the gate interval where Kondo correlations
are important, one has to calculate the Kondo temperature.
Another visualization of the Kondo region is to plot the phase of the
dot Green function
\begin{equation}
\phi=atan(Im G_{dd}/Re G_{dd}),
\end{equation}
that is known to "freeze" at the value $\pi /2$ in the Kondo regime.
The phase is plotted in Fig.3 for several temperatures, and on each
curve, we mark the point where the Kondo temperature  equals  the
temperature of the curve. The Kondo temperature was computed with
the Haldane \cite{meir05,haldane} formula. One can notice from Fig.3
that the crossing point $E_{d}\approx -0.05$ (see Fig.2) is indeed
in the Kondo regime. For higher temperatures we have checked that
the isotherms no longer pass through the crossing point as the
transport is no more governed by the Kondo physics.

In conclusion, we have developed an exact current formula easy to apply
to multiple-connected meso-systems. The advantage of the formula
is that it uses only retarded Green functions and can be applied
to any system with left-right and time reversal symmetry with
one-site interaction. In the absence of the symmetries we give an
alternative approach based on the Ng's ansatz, which nevertheless
is not sufficiently accurate for describing the correlations.
It has been found that the interplay between
the Kondo correlation  and the interference effect gives rise to
 different temperature behaviors depending on the
gate potential applied on the interacting dot, separated by a crossing
point.

{\bf Acknowledgements}. We are grateful to J.Zittartz,
 A.Rosch, L.Craco and B.R.Bu{\l}ka for helpful discussions.
 We acknowledge the financial support of Sonderforschungsbereich 608
at the  Institute of Theoretical Physics, University of Cologne,
and of the CEEX-Research Programme.


\begin{thebibliography}{10}
\bibitem{note} In particular, the Ng approximation allows to
calculate separatelly $J_{\alpha}$ and $J_{\beta}$ and we have
checked that the current conservation is fulfilled for the
considered system.
\end{thebibliography}
\end{document}